\input harvmac
\input epsf.tex
\overfullrule=0mm
\newcount\figno
\figno=0 
\def\fig#1#2#3{
\par\begingroup\parindent=0pt\leftskip=1cm\rightskip=1cm\parindent=0pt
\baselineskip=11pt
\global\advance\figno by 1
\midinsert
\epsfxsize=#3
\centerline{\epsfbox{#2}}
\vskip 12pt
{\bf Fig.\the\figno:} #1\par
\endinsert\endgroup\par
}
\def\figlabel#1{\xdef#1{\the\figno}}
\def\encadremath#1{\vbox{\hrule\hbox{\vrule\kern8pt\vbox{\kern8pt
\hbox{$\displaystyle #1$}\kern8pt}
\kern8pt\vrule}\hrule}}
\def\appendix#1#2{\global\meqno=1\global\subsecno=0\xdef\secsym{\hbox{#1.}}
\bigbreak\bigskip\noindent{\bf Appendix. #2}\message{(#1. #2)}
\writetoca{Appendix {#2}}\par\nobreak\medskip\nobreak}

\def\tvi{\vrule height 12pt depth 6pt width 0pt}
\def\tv{\tvi\vrule}


\Title{T97/151  }
{{\vbox {
\medskip
\centerline{{\bf Folding of the Triangular Lattice in the FCC Lattice}}
\centerline{{\bf with Quenched Random Spontaneous Curvature}}
}}}
\bigskip
\medskip
\centerline{S. Mori}
\centerline{\it Department of Physics, School of Science,}
\centerline{\it Kitasato University, Kitasato 1-15-1}
\centerline{\it Sagamihara, Kanagawa 228, Japan}
\medskip
\centerline{E. Guitter\footnote{}{\kern -20pt email: 
guitter@spht.saclay.cea.fr mori@sci.kitasato-u.ac.jp}}
\medskip
\centerline{ \it CEA, Service de Physique Th\'eorique de Saclay,}
\centerline{ \it F-91191 Gif sur Yvette Cedex, France}
\baselineskip=12pt
\vskip .5in
 
We study the folding of the regular two--dimensional triangular lattice
embedded in the regular three--dimensional Face Centered Cubic lattice,
in the presence of quenched random spontaneous curvature. 
We consider two types of quenched randomness: (1) a ``physical'' randomness 
arising from a prior random folding of the lattice, creating a prefered
spontaneous curvature on the bonds; (2) a simple randomness where the
spontaneous curvature is chosen at random independently on each bond.
We study the folding transitions of the two models within the hexagon 
approximation of the Cluster Variation Method. Depending on the type of 
randomness, the system shows different behaviors. 
We finally discuss a Hopfield-like model as an extension of 
the physical randomness problem to account for the case where 
several different configurations are stored in the prior pre-folding
process.

\noindent
\Date{11/97}
 
\nref\NPW{``Statistical Mechanics of Membranes and Surfaces,'' D.R. Nelson, 
T. Piran and S. Weinberg eds, Proceedings of the fifth Jerusalem Winter 
School for Theoretical Physics (World Scientific, Singapore, 1989).}
\nref\DGZ{``Fluctuating Geometries in Statistical Mechanics and Field
Theory,'' F. David, P. Ginsparg and J. Zinn-Justin eds; Les Houches
Session LXII (Elsevier Science, The Netherlands, 1996) 
(http://xxx.lanl.gov/lh94).}
\nref\MBB{M. Mutz, D. Bensimon, M. J. Brienne, 
Phys. Rev. Lett. {\bf 67} (1991) 923.}
\nref\KKN{Y. Kantor, M. Kardar and D. R. Nelson, Phys. Rev. Lett. 
{\bf 57} (1986) 791; Phys. Rev. {\bf A36}(1987) 3056.}
\nref\LG{A. E. Lobkovsky, S. Gentges, Hao Li, D. Morse and T. A. 
Witten, Science {\bf 270} (1995) 1482; E. M. Kramer and 
A. E. Lobkovsky, Phys. Rev. {\bf E53} (1996) 1465 (cond-mat/9510090); 
A. E. Lobkovsky, Phys. Rev. {\bf E53} (1996) 3750 (cond-mat/9609069).}
\nref\KJ{Y. Kantor and M.V. Jari\'c, Europhys. Lett. {\bf 11} (1990) 157.}
\nref\DIG{P. Di Francesco and E. Guitter, Europhys. Lett. {\bf 26} (1994)
455 (cond-mat/9402058).}
\nref\DIGG{P. Di Francesco and E. Guitter, Phys. Rev. E {\bf 50}
(1994) 4418 (cond-mat/9406041).}
\nref\BDGG{M. Bowick, P. Di Francesco, O. Golinelli and E. Guitter,
Nucl. Phys. {\bf B450} [FS] 463 (1995) (cond-mat/9502063).}
\nref\BDGGII{M. Bowick, P. Di Francesco, O. Golinelli and E. Guitter, 
``Discrete Folding'', Proceedings of the 4th Chia
Meeting on ``Condensed Matter and High-Energy Physics'', September
4-8, 1995 (World Scientific, Singapore) (cond-mat/9610215).}
\nref\BGGM{M. Bowick, O. Golinelli, E. Guitter and S. Mori, 
Nucl. Phys. {\bf B495} [FS] 583 (1997) (cond-mat/961105).}  
\nref\DGM{P. Di Francesco, E. Guitter and S. Mori, 
Phys. Rev. {\bf E55} (1997) 237 (cond-mat/9607077).}
\nref\MA{D. C. Mattis, Phys. Lett. {\bf A56} (1976) 421, [Erratum,
{\bf A60} (1977) 492].}
\nref\K{R. Kikuchi, J. Chem. Phys. {\bf 60} (1974) 1071.}
\nref\MO{T. Morita, Prog. Theor. Phys. {\bf 103} (1984) 103.}
\nref\An{G. An, J. Stat. Phys. {\bf 52} (1988)  727.}
\nref\CGP{E. Cirillo, G. Gonnella and A. Pelizzola, Phys. Rev. {\bf E53}
(1996) 1479 (hep-th/9507161); Phys. Rev. {\bf E53} (1996) 3253
(hep-th/9512069).}
\nref\HOP{J. J. Hopfield, Proc. Natl. Acad. Sci. USA, {\bf 79}(1982) 2554.}
\nref\AM{D. J. Amit,``Modeling Brain Function," (Cambridge University
Press, 1989) and references therein.}


The statistical properties of polymerized membranes have been extensively 
studied in the past few years [\xref\NPW,\xref\DGZ]. 
Some particular attention was paid to the role of quenched disorder
in the elasticity of the membrane, with mainly two motivations. 
The first one is to understand the mechanism of the ``wrinkling" 
transition of partially polymerized lipid vesicles [\xref\MBB].
Such membranes undergo a reversible phase transition from a high-temperature 
soft phase with strong fluctuations to a low-temperature rigid and highly 
wrinkled phase. A second motivation is, at a macroscopic level, the study of 
the statistical properties of randomly crumpled paper or, more generally, 
randomly crumpled elastic sheets [\xref\KKN,\xref\LG]. In a random crumpling 
process, creases are created, which generate random spontaneous curvature.
When iterated, the random crumpling processes can moreover cause frustration,
and the crumpled paper may then have many equally probable configurations
of minimal energy, a usual characteristics of random spin systems.  
 
Here we study a simple system of two-dimensional polymerized
object with quenched disorder. As a toy model, we consider the problem
of {\it folding} of the regular two--dimensional triangular lattice 
in the presence of random spontaneous curvature. 
Models of folding have been introduced in \KJ\ and studied in 
[\xref\DIG,\xref\DIGG]. Originally, the study was restricted to 
{\it planar folding}, i.e folding in a two--dimensional embedding space.  
A more general discrete folding model with a three--dimensional 
embedding space was then introduced and studied in [\xref\BDGG,
\xref\BDGGII,\xref\BGGM], 
describing the folding of the triangular lattice in the regular 
three-dimensional Face Centered Cubic (FCC) lattice. 
The role of disorder in the {\it planar folding} problem was analyzed 
by the authors and Di Francesco in \DGM .There, disorder was introduced 
in the form of a quenched random bending rigidity. Here we would like to 
complete our study by considering the folding in the three--dimensional FCC 
lattice in the presence of {\it quenched random spontaneous curvature}.

Before we proceed to our study, let us recall the results of \DGM\ for 
the planar folding problem with random bending rigidity. 
There, we were interested in modeling the randomness arising from 
a prior irreversible ``crumpling" of the lattice. Such a crumpling results 
in a marking of the lattice bonds with quenched creases on which folds 
are favored.
The system can then be described by a Mattis-like model \MA\ with Hamiltonian
\eqn\hammattis{ {\cal H}_{\rm Mattis} = 
-K \sum_{{\rm n.n.}(ij)} \tau_{i}\tau_{j}\sigma_i \sigma_j \ ,}
where the variables $\sigma_i=\pm 1$ describe the (up or down) normal to 
the triangle $i$ in the folded configuration and $n.n.(ij)$
means summation over nearest neighbour pairs. The disorder variables 
$\tau_{i}=\pm 1$, accounting for the prior irreversible marking, 
define a random bending rigidity $K_{ij}=K\tau_i\tau_j$, and are 
``frozen'' according to a specified probability distribution \DGM .
For the variables $\sigma$ to represent actual folded configurations 
of the lattice, the six neighboring spins on an elementary hexagon 
of the lattice, $\sigma_{i} (i=1,2,\cdots,6)$, must satisfy the ``physical'' 
constraints [\xref\KJ-\xref\DIGG]: $\sum_{i=1}^6 \sigma_i = 0 \ {\rm mod}\ 3.$
This condition in particular prevents from absorbing the disorder
in a simple change of $\sigma_{i}$ into $\sigma_{i}\tau_{i}$.
Similarly, if the corresponding disorder variables $\tau_{i}(i=1,2,\cdots,6)$  
arise from a pre--folding process, they should also obey the physical rule 
$\sum_{i=1}^6 \tau_i = 0 \ {\rm mod}\ 3 $.
This condition on the $\tau$--variables is essential for the system 
to develop a large $K$ ``frozen phase'' where the prior irreversible
folded shape is recovered. If the $\tau$ variables are free $\pm 1$ variables 
which do not satisfy the physical constraint above, the system becomes 
frustrated and the lattice remains in a disordered phase. 


We now would like to extend these results to the case of the
three-dimensional FCC folding problem. 
A folding of the triangular lattice in the FCC lattice is simply a 
mapping sending each vertex of the triangular lattice onto a vertex 
of the FCC lattice, with the requirement that neighboring vertices on 
the triangular lattice remain nearest neighbors in the FCC lattice 
[\xref\BDGG,\xref\BDGGII], i.e. belong
to the same triangular face. The FCC lattice is indeed made of octahedra 
and tetrahedra in contact by their triangular faces. Elementary triangles
of the triangular lattice are thus sent onto elementary triangular faces
of the FCC lattice. In the folded configuration, two adjacent triangles 
can form some relative angle $\theta$, with one of the four following values:
\item{(i)} $\theta=180^\circ$ {---} no fold: the triangles are side by side;
\item{(ii)} $\theta=0^\circ$ {---} complete fold: the triangles are on top
of each other;
\item{(iii)} $\theta=\arccos(1/3)\sim 71^\circ$ {---} fold with acute
angle: the two triangles lie on two adjacent faces of the same tetrahedron 
in the FCC lattice, and
\item{(iv)} $\theta=\arccos(-1/3)\sim 109^\circ$ {---} fold with obtuse
angle: the triangles lie on two adjacent faces of the same octahedron in the 
FCC lattice.  
\par 
\noindent It was shown in \BDGG\ that these four types of folds can be 
understood as the superposition of the domain walls of two $Z_2$ variables 
$\sigma=\pm 1$ and $z=\pm 1$ living on the faces of the triangular lattice. 
The relative values $\Delta\sigma\equiv\sigma_2\sigma_1$ and
$\Delta z=z_1z_2$ for two neighboring triangles indicate which type of fold 
they form, with the correspondence displayed in Table~I.  
\midinsert
$$\vbox{\offinterlineskip
\halign{\tv\quad # & \quad\tv \quad
# & \quad \tv \quad # & \quad \tv \quad # & \quad \tv #\cr
\noalign{\hrule}
\tvi $\Delta\sigma$ & $\Delta z$ & $\phantom{X}\theta$ &angle&\cr
\noalign{\hrule}
\tvi  $\phantom{-}1$ &  $\phantom{-}1$  & $180^\circ$ &no\phantom{X}fold&\cr
\tvi  $-1$ &  $\phantom{-}1$  & $\phantom{11}0^\circ$ 
&complete\phantom{X}fold&\cr
\tvi  $\phantom{-}1$ &  $-1$  & $\phantom{1}71^\circ$ 
&acute\phantom{X}fold&\cr
\tvi  $-1$ &  $-1$ & $109^\circ$  &obtuse\phantom{X}fold&\cr
\noalign{\hrule} }} $$
\par\begingroup\parindent=0pt\leftskip=1cm\rightskip=1cm\parindent=0pt
\baselineskip=11pt {\bf Table I:} The relative folding state of two
neighboring triangles according to the relative values
$\Delta\sigma$ and $\Delta z$.  
\par 
\endgroup\par 
\endinsert
\noindent In order to describe an actual allowed folded state, the 
$\sigma$ and $z$ variables are subject to two basic 
folding rules involving the values $\sigma_i$ and $z_i$ ($i=1,\cdots,6$) 
on the six neighboring triangles forming an elementary hexagon in the
lattice. For each hexagon, the variables $\sigma$ must satisfy the
following first folding rule:
\eqn\firstrule{\sum_{i=1}^6 \sigma_i = 0 \ {\rm mod}\ 3 .}
This rule is identical to the rule of planar folding \DIG\ although
its interpretation here is slightly different [\xref\BDGG,\xref\BDGGII]. 
A second basic folding rule 
involves both the $z$ and $\sigma$ variables and reads: 
\eqn\secondrule{\prod_{i\in I(c)}z_iz_{i+1}=1 \quad{\rm for}\ c=0,1,2 \quad ;
\quad I(c)=\{i\ :\
\sum_{k=1}^i \sigma_k = c \ {\rm mod}\ 3 \}.} 
With the ``physical'' constraints \firstrule\ and \secondrule , one finds 
exactly 96 
possible hexagonal configurations for the six triangles surrounding any of the vertices 
of the triangular lattice. Note that the planar folding problem can be recovered 
by freezing the $z$ variable to $z_i=+1$ globally for all triangles. One is then
left with exactly 11 possible hexagonal configurations.

In the absence of disorder, the folding energy is simply 
$E_{\rm pure}=-K\ \cos (\theta )$ 
per lattice bond, with $K$ the bending rigidity parameter. 
In terms of the variables $\sigma_1,\sigma_2$ and $z_1,z_2$ of the two triangles 
forming the fold, the folding energy simply reads: 
\eqn\ener{E_{\rm pure}=-{K\over 3} \sigma_1\sigma_2 \,(1+2z_1z_2) .}  
The total folding energy is the sum of all elementary folding energies for
all the bonds of the triangular lattice. 

Disorder can be put in the model by introducing quenched disorder
variables $\tau_i$ and $w_i$ describing the pre--folded state created
by the irreversible crumpling process. The domain walls for 
the variables $\tau$ and $w$ encode the four possible types of created creases 
with angle $\psi=180^\circ$, $0^\circ$, $71^\circ$ or $109^\circ$ 
according to a table similar to Table~I.
Of course, in order to describe an actual pre--folded state, the 
$\tau$ and $w$ variables are subject to two local folding rules
similar to \firstrule\ and \secondrule .
The presence of random creases directly leads to a random spontaneous curvature
in the system, encoded in the angle $\psi$ of the crease. Given this angle,
the energy becomes minimum when the angle $\theta$
of the fold is such that $\theta=\psi$.  
We shall thus consider the following bending energy:
\eqn\rbend{E=-K\ \cos (\theta-\psi ) ,}
where $K$ measures the strength of this bending energy
and where the quenched random variable $\psi$ describes the quenched
random spontaneous curvature in the system.
As mentioned above, the variables $\theta$ and $\psi$ take four values,
leading to 16 possible fold/crease configurations. 
We  would like to express the energy \rbend\ in terms of the spin variables
$\sigma_i$, $z_i$, $\tau_i$ and $w_i$ ($i=1,2$) on the two neighboring triangles
on each side of the fold, as we did in \ener\ for the pure case without
disorder. The relative values $\Delta \sigma \equiv \sigma_1\sigma_2$ and 
$\Delta z\equiv z_1z_2$ on one hand and those of $\Delta \tau \equiv 
\tau_{1}\tau_{2}$ and $\Delta w \equiv w_{1}w_{2}$ on the other hand
fix the angles $\theta$ and $\psi$ and thus the energy \rbend . 

We can make use of the symmetry $\theta \leftrightarrow \psi$
in \rbend\ (i.e. $(\Delta\sigma,\Delta z)\leftrightarrow (\Delta\tau,
\Delta w)$) and of the symmetry $(\theta ,\psi)\leftrightarrow
(180^\circ-\theta,180^\circ-\psi)$ (i.e. $(\Delta\sigma,\Delta\tau)
\leftrightarrow -(\Delta\sigma,\Delta\tau)$) to ensure that 
the bending energy has the two {\it independent} symmetries
$\Delta \tau \leftrightarrow \Delta \sigma$ and
$\Delta w \leftrightarrow \Delta z$, and is even in 
$(\Delta\sigma+\Delta\tau)/2$, leading to the general form (for $Z_2$
variables):
\eqn\gbend{\eqalign{{E\over K} & =e+a{(\Delta z + \Delta w)\over 2} 
+b_{1}(\Delta \sigma \Delta \tau)+b_{2}(\Delta z \Delta w)\cr
&\ \ +c(\Delta \sigma \Delta \tau){(\Delta z + \Delta w)\over 2}
+d (\Delta \sigma \Delta \tau)(\Delta z \Delta w)\cr }
} 
involving 6 constant coefficients $e,a,b_{1},b_{2},c,d$ 
to be determined hereafter. These coefficients are simply obtained 
from the values of the 6 independent (i.e. not related by the above
symmetries) pairs of folding and disorder 
configurations given in table~II.

%
\midinsert
$$\vbox{\offinterlineskip
\halign{\tv\quad # & \quad\tv \quad
# & \quad \tv \quad  # & \quad \tv #\cr
\noalign{\hrule}
\tvi Fold Config. & Disorder\phantom{X}Config. & Bending\phantom{X}Energy/$K$ 
&\cr \noalign{\hrule}
\tvi  No\phantom{X}fold &  No\phantom{X}Crease &\phantom{1234567/}-1&\cr
\tvi  No\phantom{X}fold &  Acute\phantom{X}Crease  &\phantom{123456-}1/3&\cr
\tvi  No\phantom{X}fold &  Complete\phantom{X}Crease &\phantom{1234567/-}1&\cr
\tvi  No\phantom{X}fold &  Obtuse \phantom{X}Crease &\phantom{123456}-1/3&\cr
\tvi  Acute\phantom{X}Fold &  Acute\phantom{X}Crease &\phantom{1234567/}-1&\cr
\tvi  Acute\phantom{X}Fold &  Obtuse\phantom{X}Crease &\phantom{123456}-7/9&\cr
\noalign{\hrule} }} $$
\par\begingroup\parindent=0pt\leftskip=1cm\rightskip=1cm\parindent=0pt
\baselineskip=11pt {\bf Table II:} The folding and crease state 
of two neighboring triangles and the corresponding  bending energy$/K$ for
the 6 independent configurations.  
\par 
\endgroup\par 
\endinsert
\noindent {}From this table, we determine the bending energy to be:
\eqn\tbend{\eqalign{E&= {K\over 9}\Big[-2+4{(\Delta z + \Delta w)\over 2}
-(\Delta \sigma \Delta \tau)
-2(\Delta z \Delta w)  \cr
&\  -4(\Delta \sigma \Delta \tau){(\Delta z + \Delta w)\over 2}
-4 (\Delta \sigma \Delta \tau )(\Delta z \Delta w)\Big]\cr 
&=-{K\over 9}[\Delta \sigma(1+2\Delta z) \Delta \tau(1+2\Delta w)
+2(1-\Delta z)(1-\Delta w)]\ .}}
The total bending energy is again the sum of all elementary folding 
energies for all links of the triangular lattice.
It is interesting to check several limiting cases of the above formula.
The pure folding problem, without disorder, can be recovered by constraining 
the disorder variables according to $\Delta \tau =\Delta w =1$. The bending 
energy \tbend\ reduces to $E=-K\Delta \sigma(1+2\Delta z)/3$, i.e to Eq.\ener . 
The case of planar folding with disorder is obtained by setting
$\Delta z =\Delta w =1$ in \tbend\ and the bending energy reduces then 
to $E=-K\Delta \tau \Delta \sigma$, as in Eq.\hammattis . 

In addition to the bending energy \tbend , we also introduce  
an external field $H_{r}$ associated with the variable $(\sigma\tau+wz)$,
which is a rough measure of how close the folding configuration
$(\sigma,z)$ is from the disorder configuration $(\tau,w)$.
This definition of the external field is not canonical and many
other definitions are equally acceptable.
The main motivation for introducing this external field is technical,
i.e. the necessity to first prepare a solution with explicit broken symmetry 
to be able to eventually reach a solution with spontaneous symmetry 
breaking by tuning $H_r$ to zero.
Our total Hamiltonian is thus given by:
\eqn\tham{{\cal H}=-{K\over 9}\sum_{{\rm n.n.} (ij)}
[\tau_i\tau_j(1+2 w_iw_j)\sigma_i\sigma_j(1+2 z_iz_j)
+2(1-w_iw_j)(1-z_iz_j)]-H_r\sum_i(\sigma_i\tau_i+z_iw_i)}
To analyze the properties of the system, we will consider
different order parameters. In view of what we know about
the pure system without disorder, it is useful to
divide the original triangular lattice into the two subsets A and B 
made of all triangles pointing up and all triangles pointing 
down respectively in the original flat triangular lattice.
This in particular allows us to define ``staggered'' average values. 
We will be interested primarily in the four following average values:
\eqn\OP{\eqalign{
&S_{\rm A}\equiv \overline{\langle \sigma \rangle_{\rm A}}\ , 
\cr &Z_{\rm A}\equiv \overline{\langle z \rangle_{\rm A}}\ ,
\cr &F_{1}\equiv \overline{\langle \sigma \tau \rangle}\ ,\cr
&F_{2}\equiv \overline{\langle z w \rangle}\ ,\cr}}
where the brackets denote the average over the configurations
at fixed disorder for an arbitrary given triangle (taken moreover
in the subset A when we add the index A to the brackets) and the over-line
denotes the average over the quenched disorder. One has of course 
$-1\le S_{A},Z_{A},F_{1},F_{2}\le 1$. 
Non-zero values of the ``frozen'' order parameters $F_{1}$ and $F_{2}$
indicate that the membrane is trapped in the configuration given by the 
disorder variables $\tau$ and $w$.
The reason why we restrict ourselves to the subset A in $S_{\rm A}$
and $Z_{\rm A}$ is that, for the  ``pure'' system, that is the model
without disorder, and at $K=0$, it was established that the lattice 
is found in a phase where $S_{\rm A}=-S_{\rm B}\sim 0.874560$ 
and $Z_{A}=Z_{B}=0$. This can be interpreted as a strong preference 
for the lattice to wrap on octahedra in the FCC lattice.
In this phase, clearly the full average value 
$S\equiv (S_{\rm A}+S_{\rm B})/2=0$ is not a good quantity and the correct
order parameter is $S_{\rm st}\equiv (S_{\rm A}-S_{\rm B})/2$, or
more simply $S_{\rm A}$ itself. 
 
We now come to the question of the precise form of the probability 
distribution for the quenched disorder variables $\tau$ and $w$.  
As has been discussed previously, these disorder variables should obey 
the two folding constraints in order to correspond to some prior folding of 
the lattice. As in our previous work in \DGM , we can take advantage of 
the solution of the pure system and simply assume that the 
disorder distribution is described by a particular equilibrium
distribution of this pure system. Since there is no physical reason 
to introduce an energy scale in the distribution of the disorder 
variables and because we want to treat as equiprobable all pre-folded 
configurations, the natural choice is to take the distribution of the 
pure system at $K=0$. As we just mentioned, the triangular lattice is then 
in a ``octahedrally" folded phase. The disorder configuration will thus
also have the same nature, i.e. $\overline{\tau_{\rm st}}=0.874560$.  
This means that the disorder configuration is dominated by obtuse 
and complete creases.
We will refer to the disordered model with this particular probability 
distribution as Model 1. For comparison, we will also study a model 
without the physical constraints on the $\tau$ and $w$ variables, i.e
a case were the two random variables take $\pm 1$ values
with equal probability, independently on each triangle. All 
the $(2\times2)^{6}=4096$ hexagonal disorder configurations are then
possible and equiprobable. We shall call refer to this second model Model 2 
and will compare its behavior with that of Model 1.

To analyze the properties of both Models 1 and 2, we rely on 
the same method we used in \DGM\ for the planar case, i.e.
the Cluster Variation Method (CVM) in its hexagonal approximation.
The CVM uses a variational principle on the free energy of the system
together with a suitable truncation of the cumulant expansion of the 
entropy at the level of some maximal clusters (here the hexagons)
[\xref\K-\xref\An]. It is applicable to the statistics of both 
pure systems without disorder and to systems with quenched 
random disorder. In this case, it allows in particular to evaluate 
average values such as \OP . This method has been first used for the
planar and FCC folding problems in \CGP .
Although it is only an approximation since the entropy
is evaluated from a truncation of its cumulant expansion,
it has been shown in the most simple cases of folding problems 
that its results compare extremely well with exact predictions \CGP .
For more complex cases where no exact results exist, we still
believe that this method allows for reliable and accurate predictions,
as far as the nature of the phases and of the transitions are concerned.
We refer to our previous work \DGM\ for a detailed description of 
this method and its implementation for disordered systems.

\fig{The free energy and the order parameters $S_{\rm A}$ (dashed),
$Z_{\rm A}$ (solid), $F_{1,C}$ (dashed) and $F_{2}$ (solid) 
as a function of the bending energy $K$ and the external field $H_{r}$. 
These results are for Model 1, i.e. with disorder variables satisfying
physical constrains.}{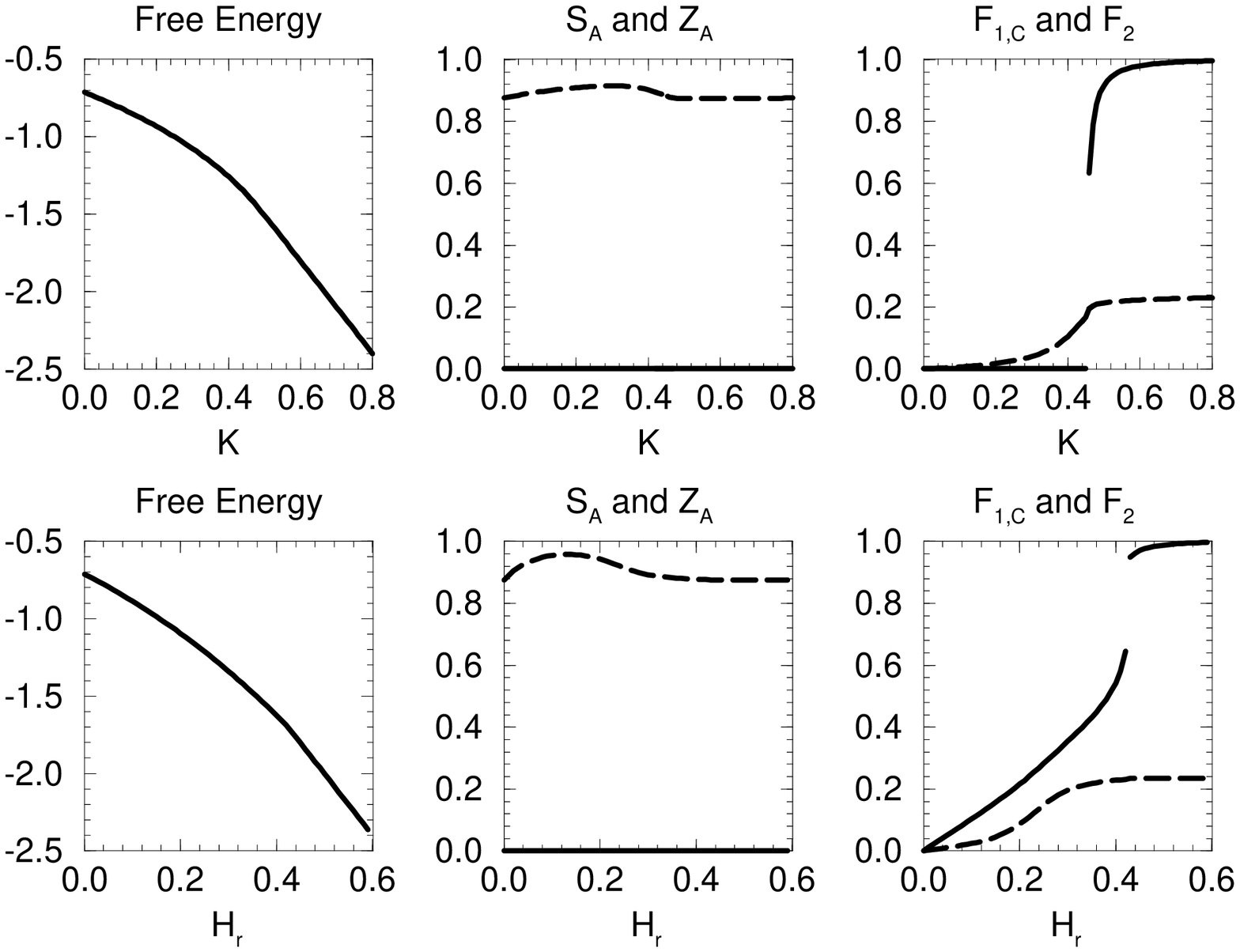}{10.truecm} \figlabel\octa

We now present the results of our CVM analysis.
Our main result is, as for the planar system, the dependence of the 
existence of a frozen phase $F_{i}\neq 0 \ (i=1,2)$ on the type of 
disorder. In the case of Model 1, a discontinuous transition occurs
at $K=K_{F3}\sim 0.44$ from a low $K$ octahedrally folded phase to
a large $K$ frozen phase. Model 2 does not develop such a phase
and the lattice is always found in the octahedrally folded state 
(this was checked up to $K=10.0$).
We have also looked at the effect of the external field $H_{r}$
alone.
In this case, both systems show a discontinuous transition to a
partially frozen phase $0<F_{i}<1$, with however very different
resulting states. In the frozen phase of the Model 1, the $F_{i}$ almost 
saturate to $1$ and the lattice is thus completely trapped in 
the configuration specified by the disorder variables. 
In the case of Model 2, the values of $F_{i}$ are much smaller, which 
means that the degree of freezing is far from complete. 
Such a different character comes from the frustration appearing 
in the system for Model 2, a result similar to what we found in the planar 
folding model \DGM. 
The external field tries to put the lattice
configuration into a disorder configuration which is in general not accessible
due to the folding constraints on the $\sigma$ and $z$ variables.  
In other words, there is not a unique ground state for 
arbitrary disorder variables.

\fig{The free energy and the order parameters $S_{\rm A}$ (dashed),
$Z_{\rm A}$ (solid), $F_{1,C}=F_{1}$ (dashed) and $F_{2}$ (solid) 
as a function of the bending energy $K$ and the
external field $H_{r}$. These results are for Model 2, i.e. with disorder
variables completely random.}{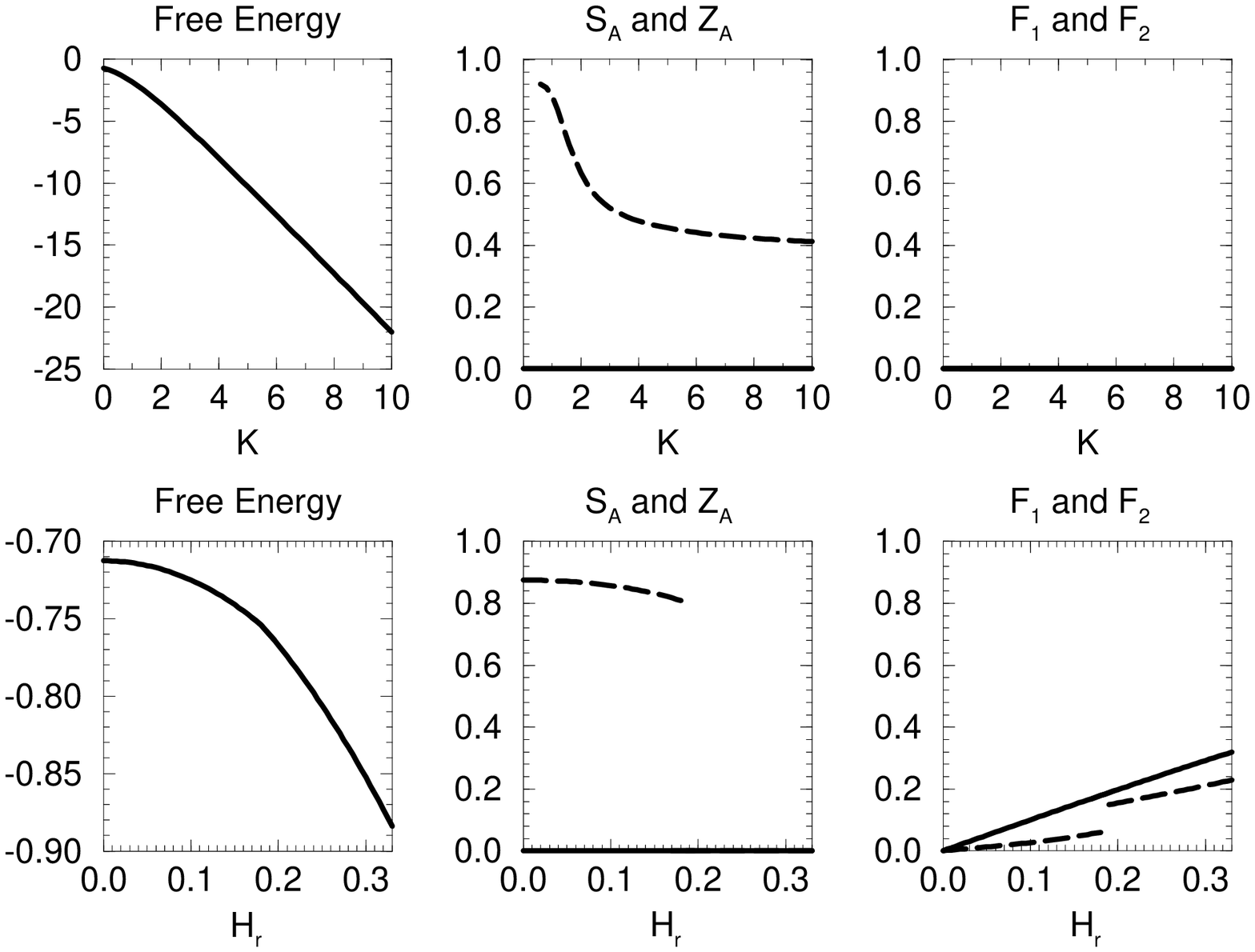}{10.truecm} \figlabel\random

In Fig.\octa\ and Fig.\random , we show the behaviors of the free energy
and the different order parameters $S_{\rm A}$, $Z_{\rm A}$, 
$F_{1,C}\equiv F_{1}- S_{A}\overline{\tau_{A}}$ and $F_{2}$ 
for both the Model 1 and the Model 2 as a function of the bending energy $K$ 
and the external field $H_{r}$. 
We display $F_{1,C}$ instead of $F_{1}$, because in the case of Model 1, 
$F_{1}$ takes non-zero values even at $(K,H_{r})=(0,0)$. Indeed, when $K=H_r=0$,
the disorder variables $\tau$ and $w$ are decoupled from the physical 
degrees of freedom $\sigma$ and $z$. {}From the definition of $F_{1}$, its value is
then simply given by the product of the value $S_{\rm A}$ of the pure
system at $K=0$ and the quantity $\overline{\tau_{\rm A}}$, both equal
to $0.874560$, leading to $F_{1}=(0.874560)^{2}\sim 0.765$. This non-zero value
is subtracted if we use the ``connected part'' $F_{1,C}$ instead.

In the case of Model 1, by increasing the bending rigidity $K$, 
the systems undergoes a phase transition 
from the octahedrally folded phase ($S_{\rm A}\neq 0$, $Z_{\rm A}=0$,
$F_{1,C}\neq 0$ and $F_{2}=0$) 
to the frozen phase ($S_{\rm A}\sim \overline{\tau_{\rm A}}
=0.874560$, $Z_{\rm A}=0$, 
$F_{1,C} \sim 1.-(0.874560)^{2}\sim 0.235$, $F_{2}\sim 1$). 
The values of $F_{i}$ saturate to $1$, meaning that the system is almost completely 
trapped in the configuration specified by the disorder variables. 
The external field also induces a phase transition. In this case, 
a weak field already creates a non-zero value of $F_{2}$ and the lattice 
is thus in a weakly frozen phase. A discontinuous transition occurs to a 
highly frozen phase at with strong freezing $F_{i}\sim 1\  (i=1,2)$.  

In the case of Model 2, increasing the bending rigidity does not give
rise to a frozen phase (in the range $K <10.0$ at least were we performed 
our analysis). The octahedral order $S_{\rm A}\neq 0$ persists for all
$K$ and the triangular lattice remains in the octahedrally folded phase. 
With the external field $H_{r}$, the octahedral order disappears abruptly 
and a partially frozen phase $F_{i}\neq 0\ (i=1,2)$ appears at $H_{r,C}=0.18$. 
Still, the values of $F_1$ and $F_2$ are small in this phase compared with 
the large $H_{r}$ frozen phase of Model 1. 

\bigskip

In the present work and in our previous work \DGM , we have studied 
triangular lattice folding models with random bending rigidity and
random spontaneous curvature. These models can be seen as toy models
for crumpled paper in the sense that they combine the effect of disorder
and of geometrical metric constraints on a geometrical two--dimensional 
object.  We have in mind the situation where the paper is  
crumpled only once and the system thus does not have frustration
and can recover this crumpled state.
A natural question is now what happens if the paper is crumpled several times.
In this case, we expect that the lattice should store several
creases configurations. Let us discuss now how this can be implemented
in our model.
For simplicity we return to the {\it planar} folding problem
with random bending rigidity (see Eq.\hammattis).
We can imagine that the lattice, after $p$ crumpling processes, 
stores $p$ different configurations, 
which we denote by $\tau_{i}^{\mu} 
(\mu=1,\cdots,p)$. 
Then one possible choice is the following bending energy, 
written by analogy with the Hopfield model [\xref\HOP,\xref\AM]:
\eqn\khop{ K_{ij}={K\over p} \sum_{\mu=1}^{p}\tau_{i}^{\mu}\tau_{j}^{\mu}.}
We note however two differences with the usual Hopfield model.
In our problem, the spin variables $\sigma_{i}$ and the 
disorder random variables $\tau_{i}^{\mu}$ correspond to folded
configurations of the lattice, and must satisfy the planar folding 
constraint (Eq.\firstrule). Moreover,
in the Hopfield model, the interaction
is long-ranged $(\sum_{(i,j)})$, while in our model, 
 it is simply short-ranged $(\sum_{{\rm n.n.} (ij)})$. 

\fig{An elementary hexagon with two different 
memories $\tau^{1}$ and $\tau^{2}$. 
We show on the r.h.s. the corresponding values of $K_{ij}$ according
to (a) Eq.(10) and (b) Eq.(11). 
}{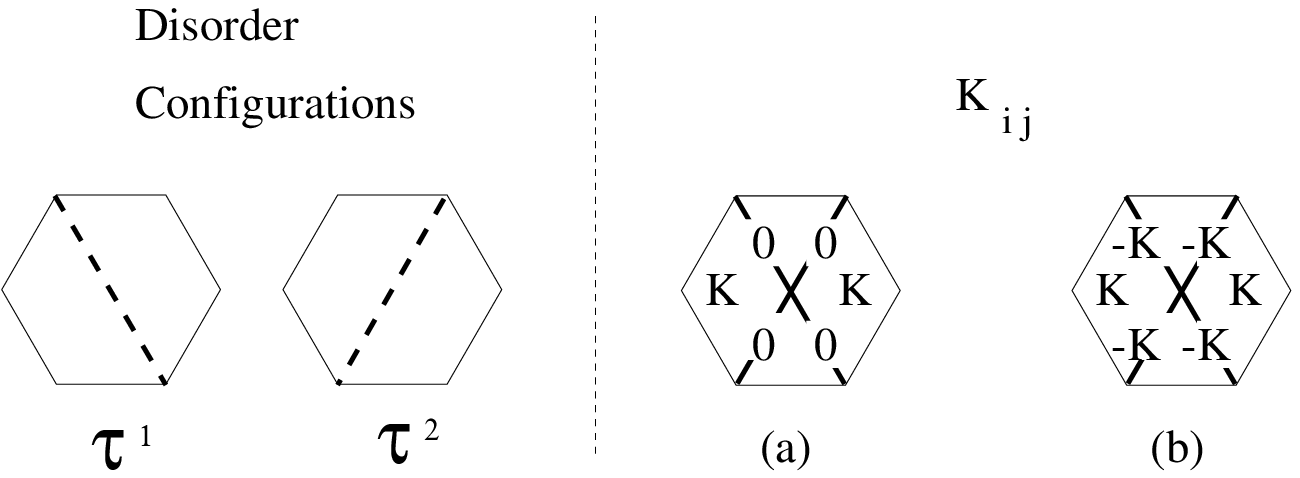}{10.truecm} \figlabel\frus 

\noindent The choice \khop\ for the bending energy is however somewhat unsatisfactory.
To see why, we consider an elementary hexagon which tries to store the 2 disorder 
configurations of Fig.\frus . With the definition of Eq.\khop, 
the bending rigidity $K_{ij}$ at the creases becomes zero (see Fig.\frus-a), 
a result which
is somewhat unrealistic. We would instead expect that,
if one folds the hexagon as in the Fig.\frus,  
the bending rigidity becomes negative at all the creases and remains positive 
where there is no crease (see Fig.\frus-b). 
A more natural choice for the bending energy is then
\eqn\khops{ K_{ij}={K} {\rm min}[\tau_{i}^{\mu}\tau_{j}^{\mu}(\mu=1,\cdots,p)].}
This choice corresponds to a complete irreversibility 
of the process of marking creases.
%
%

\fig{An elementary hexagon with unphysical disorder and
the corresponding lowest energy configurations satisfying
the folding constraint.}{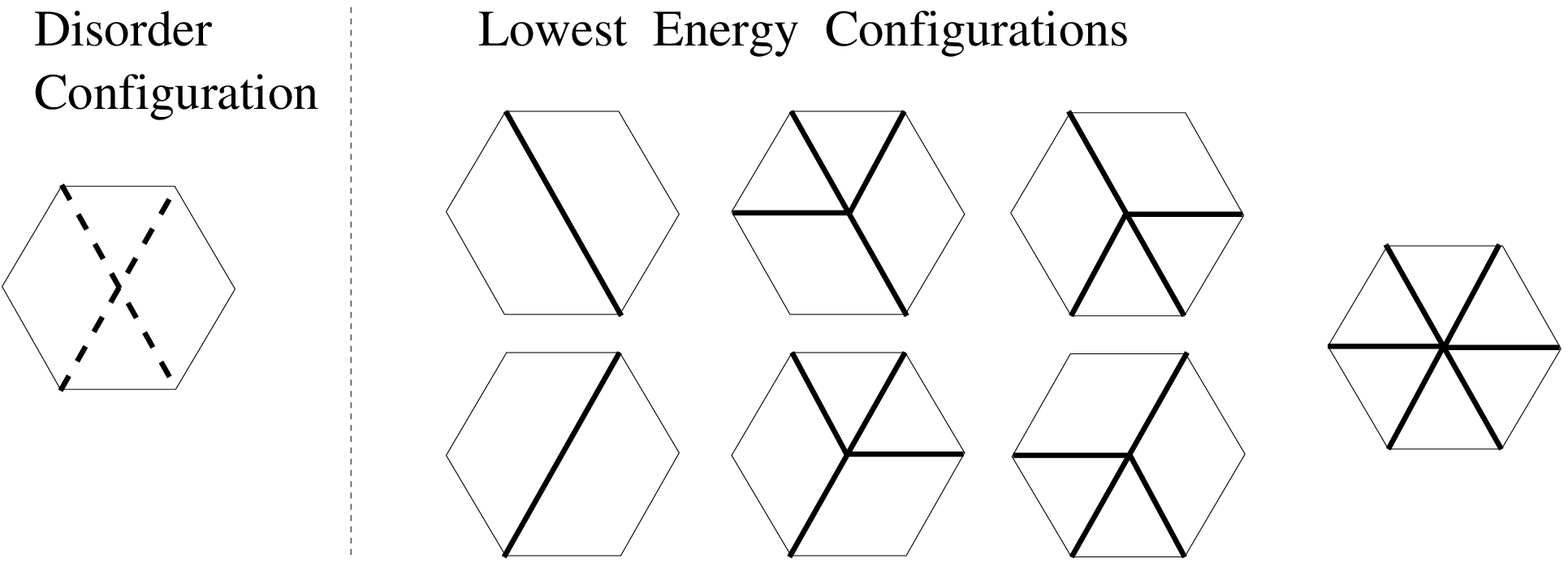}{10.truecm} \figlabel\fruss

With such a disorder, we now find for the case of Fig.\frus\ seven
ground states in competition (Fig.\fruss). 
This situation is the exactly similar to the case 
of unphysical disorder, where the $\tau$ variables do not satisfy 
the folding constraint. 
In our previous work, we have shown that the strong 
frustration caused by such an unphysical disorder prevents the
system from having a frozen phase at large $K$.
Based on this result, we could conclude that, in the presence of 
many different stored configurations, the system 
remains always disordered. On the other hand,
as we increase the number of irreversible crumplings, the number of 
creases becomes large and the system becomes almost identical to 
a {\it pure} model with negative bending rigidity. Such a pure system 
is known to be at large $K$ in a so-called ``piled up" ordered phase.
We can assume that one develops such an order if the number $p$ of 
stored configuration exceeds a critical value $p_{c2}$.
Finally, from the properties of the Hopfield model, 
on can also imagine the system to be in the so-called spin-glass phase. 

{}From the above discussion, we can propose the following conjecture 
on the statistical properties of the above model at $K=\infty$, which  
is of particular importance because crumpled paper can be considered 
as an infinite elastic constant limit $K\rightarrow \infty$ 
(or $T=0$) of the lattice system. 
The triangular lattice with $p$ stored configurations could be in the 
retrieval phase for very small $p<p_{c1}$, where the system 
can almost store and retrieve $p$ patterns.
For $p_{c1}<p<p_{c2}$, the system would be in a spin-glass phase 
or in a disordered phase.
Finally, for $p>p_{c2}$, the system would be in the piled up phase. 
The values of $p_{c1}$ and $p_{c2}$ are expected to be very small, because 
the interaction is short-ranged and the probability that each elementary 
hexagon has two or three creases is very high even with one stored 
configuration. 
 
Finally, we also expect that the many crumpled configurations which are 
caused by successive crumplings of the paper are not independent from one 
another and we can suppose that they look quite similar. This weakens 
the frustration, which might keep the paper in the retrieval phase for 
larger $p$, i.e. increase the value of $p_{c1}$ and $p_{c2}$. In fact, 
it was also reported experimentally \LG\ that the paper has a good memory 
of the previous crumpled configuration. 
Even in the retrieval phase, the energy landscape in the phase space of 
the system can be very complex. 
It is this complex nature which might appear as the universal power 
law of the noise emitted from the crumpled elastic sheets \LG .
The thermodynamics of the above model are left for a future 
study. In addition, more realistic bending energy could be 
considered. 
To conclude, we can say that the problem of multiple stored configurations is
still very open.

\bigskip

\noindent{\bf Acknowledgments}

We thank Dr. M. Bowick and Dr. Y. Ozeki for useful 
discussions.

\listrefs

\end